\documentclass[twocolumn,floatfix,superscriptaddress,a4paper,showpacs,showkeys,nofootinbib,reprint,prc]{revtex4-1}
\usepackage{epsfig}
\usepackage{latexsym}
\usepackage[colorlinks=true,linktocpage=true,linkcolor=blue,citecolor=blue,allcolors=blue]{hyperref}
\usepackage{url}
\usepackage[utf8]{inputenc}
\usepackage{enumerate}
\usepackage{color}
\usepackage{xcolor}
\usepackage{amsmath,amssymb}

\usepackage[english]{babel}
\usepackage{hyperref}
\usepackage{url}
\usepackage{needspace}
\hypersetup{
   colorlinks=true, 
  linktoc=all,     
  linkcolor=blue,  
}

\def\be{\begin{equation}}
\def\ee{\end{equation}}
\def\bearr{\begin{eqnarray}}
\def\eearr{\end{eqnarray}}

\begin{document}

\title{The time evolution of light nuclei cumulants and ratios with a first-order phase transition in the UrQMD transport model}

\author{Thiranat Bumnedpan}
\affiliation{Frankfurt Institute for Advanced Studies (FIAS), Ruth-Moufang-Str. 1, D-60438 Frankfurt am Main, Germany}
\affiliation{Center of Excellence in High Energy Physics \& Astrophysics, School of Physics, Suranaree University of Technology, University Avenue 111, Nakhon Ratchasima 30000, Thailand}

\author{Jan Steinheimer}
\affiliation{GSI Helmholtzzentrum f\"ur Schwerionenforschung GmbH, Planckstr. 1, D-64291 Darmstadt, Germany}
\affiliation{Frankfurt Institute for Advanced Studies (FIAS), Ruth-Moufang-Str. 1, D-60438 Frankfurt am Main, Germany}

\author{Tom Reichert}
\affiliation{Frankfurt Institute for Advanced Studies (FIAS), Ruth-Moufang-Str. 1, D-60438 Frankfurt am Main, Germany}
\affiliation{Institut f\"{u}r Theoretische Physik, Goethe-Universit\"{a}t Frankfurt, Max-von-Laue-Str. 1, D-60438 Frankfurt am Main, Germany}
\affiliation{Helmholtz Research Academy Hesse for FAIR (HFHF), GSI Helmholtzzentrum f\"ur Schwerionenforschung GmbH, Campus Frankfurt, Max-von-Laue-Str. 12, 60438 Frankfurt am Main, Germany}

\author{Christoph Herold}
\affiliation{Center of Excellence in High Energy Physics \& Astrophysics, School of Physics, Suranaree University of Technology, University Avenue 111, Nakhon Ratchasima 30000, Thailand}

\author{Ayut Limphirat}
\affiliation{Center of Excellence in High Energy Physics \& Astrophysics, School of Physics, Suranaree University of Technology, University Avenue 111, Nakhon Ratchasima 30000, Thailand}

\author{Marcus Bleicher}
\affiliation{Institut f\"{u}r Theoretische Physik, Goethe-Universit\"{a}t Frankfurt, Max-von-Laue-Str. 1, D-60438 Frankfurt am Main, Germany}
\affiliation{GSI Helmholtzzentrum f\"ur Schwerionenforschung GmbH, Planckstr. 1, D-64291 Darmstadt, Germany}
\affiliation{Helmholtz Research Academy Hesse for FAIR (HFHF), GSI Helmholtzzentrum f\"ur Schwerionenforschung GmbH, Campus Frankfurt, Max-von-Laue-Str. 12, 60438 Frankfurt am Main, Germany}

\date{\today}

\begin{abstract}
The UrQMD model with a density dependent equation of state, including a first-order phase transition, is used to study the time dependence of baryon number and proton number susceptibilities up to third order in heavy ion reactions of $E_{\mathrm{lab}}=2-3 A$\,GeV. A significant deviation from the Gaussian fluctuations of the baryon number fluctuation in coordinate space is observed. The proton number fluctuations are always suppressed as they constitute only a small fraction of the total baryon number during the dense phase of the collision. It is found that the only measurable, but small, signal would be an enhancement of the third order (or higher) proton cumulant in a finite rapidity window $\Delta y$ that is larger than one unit of rapidity. In addition, it is found that the coordinate fluctuations will lead to an enhancement of cluster production due to the correlations in coordinate space. However, this enhancement is small and mainly occurs during the dense part of the collision before the system actually freezes out.  
\end{abstract}

\maketitle

\section{Introduction}

The possibility of a first-order phase transition in the phase diagram of high density QCD has been at the center of decades-spanning experimental programs at several large collider facilities. As direct calculations with lattice QCD are unfeasible at finite chemical potentials one has to rely on experimental signals in relativistic heavy ion collisions \cite{NA49:1999myq,STAR:2002eio,PHENIX:2003nhg,Gazdzicki:2008kk,ALICE:2008ngc,HADES:2009aat} as well as neutron star properties \cite{Miller:2019cac,Riley:2019yda,Miller:2021qha,Riley:2021pdl} and binary neutron star mergers \cite{LIGOScientific:2018cki,LIGOScientific:2020aai,LIGOScientific:2020zkf} and supernovae observations \cite{Jakobus:2023fru}. 
To verify the existence of such a phase transition in heavy ion reactions, one has to compare experimental data with simulations that can incorporate features of such a transition. These models range from one- and multi-fluid \cite{Hama:2004rr,Bass:2000ib,Hirano:2001eu,Kolb:2003dz,Petersen:2008dd,Werner:2010aa,Gale:2013da,Shen:2014vra,Herold:2022laa,Lin:2004en,Cimerman:2023hjw} simulations to microscopic transport models \cite{Bass:1998ca, Bleicher:1999xi,Li:2016uvu,Petersen:2018jag,Steinheimer:2018rnd}. Besides the collective flow, which is sensitive to the equation of state \cite{Rischke:1995pe,Stoecker:2004qu,Brachmann:1999xt,Brachmann:1999mp,Steinheimer:2022gqb,Oliinychenko:2022uvy}, correlation and fluctuation observables have been proposed which may be more sensitive to the effects of a phase transition and critical phenomena \cite{Karsch:2016yzt,Bazavov:2020bjn,Isserstedt:2019pgx,Skokov:2010uh,Almasi:2017bhq,Wen:2018nkn,Steinheimer:2012gc}.

One remaining question here is whether the correlations and fluctuations, which will develop in coordinate space, can be measured in the final momentum space distributions and correlations \cite{Savchuk:2022msa} which seem to depend on sufficiently strong flow that cannot be obtained at the low beam energies under consideration here \cite{Kuznietsov:2024xyn}. A recent work suggested that the study of light nuclear cluster production would be able to unravel the correlations in coordinate space, as their production probability increases if the constituent nucleons are closer in coordinate space \cite{Sun:2020uoj,Sun:2020pjz}. 

The purpose of our study is twofold. First, we present a study for heavy ion reactions in the presence of a strong first-order phase transition at finite density within the microscopic transport model UrQMD. To confirm the effect of the phase transition on coordinate space clumping we calculate the second and third order cumulants of the baryon number and different specific baryons during the time evolution in coordinate and momentum space. This is the first time the third order cumulant is calculated within a realistic microscopic simulation with a first-order phase transition. Then the formation probability of light nuclear clusters will be simulated with and without the phase transition and the quantitative effect of the phase transition on this observable will be studied.

\section{UrQMD and the EoS }

For this study we will employ the UrQMD model which allows us to include any density dependent EoS in the QMD part of the evolution \cite{OmanaKuttan:2022the}. The UrQMD model is based on a cascade part which describes the covariant propagation and scattering of an extensive list of hadrons based on known scattering cross sections. The imaginary part of the scattering includes resonance excitations and decay as well as string excitations \cite{Bass:1998ca, Bleicher:1999xi}. 
In addition to the cascade part, the model also includes a quantum molecular dynamics part, based on the non-relativistic equations of motion of QMD \cite{Aichelin:1986wa}:

\begin{eqnarray}\label{motion}
\dot{\textbf{r}}_{i}=\frac{\partial  \langle H  \rangle}{\partial\textbf{p}_{i}},
\quad \dot{\textbf{p}}_{i}=-\frac{\partial  \langle H \rangle}{\partial \textbf{r}_{i}},
\end{eqnarray}

where {\it $\langle H \rangle$} is the total Hamiltonian function of the system. The Hamiltonian of each baryon, $H_i=E^{\mathrm{kin}}_i + V_i$, comprises the kinetic energy and the mean field potential energy $V_i=E_{\mathrm{field}}/A$ of the baryon $i$. 

In previous studies it was shown how different density dependent equations of state can be incorporated in the UrQMD model \cite{OmanaKuttan:2022the,Steinheimer:2022gqb} including those with a strong first-order phase transition \cite{Savchuk:2022msa,Li:2022iil}. In the present setup we will use one such EoS that was presented in \cite{Savchuk:2022aev}. We refer to that paper for details on the construction of the EoS and will only introduct the basics of the model in the following.

The results presented in this paper will be based on two different equations of state. The first, called \textit{CMF} or \textit{crossover EoS}, is based on the current parametrization of the chiral mean field model \cite{Motornenko:2019arp} and resembles a smooth crossover transition at finite density. In addition we will compare this \textit{crossover EoS} to a \textit{phase transition EoS} which was constructed in \cite{Savchuk:2022aev}. This phase transition includes an unstable phase between densities of 2.5 and 4 times nuclear saturation density $n_0$. As shown in \cite{Savchuk:2022aev,Savchuk:2022msa,Li:2022iil}, with this EoS the phase transition can be reached already for central AuAu collisions at beam energies above $E_{\mathrm lab}> 1.5 A$\,GeV.  

The instabilities associated with this phase transition will lead to enhanced fluctuations of the baryon number during the evolution and the multiplicity distributions of baryons in a fixed spatial volume can be described by a bimodal distribution with two coexisting phases. 

In the following we will investigate the behavior of the second and third order cumulants which can provide additional insights into the role of the phase transition during the evolution. 

\subsection{Susceptibilities}
The baryon number fluctuations which can occur due to the phase separation during a non-equilibrium first-order phase transition can be studies by the cumulants of the multiplicity distributions in a given spatial volume. 
We will use the cumulants $K$ for this study, which can be related to the factorial cumulants \cite{Bzdak:2018axe,Kitazawa:2017ljq} used in other studies, and essentially contain the same information. The first three cumulants are defined as:

\begin{eqnarray}
\label{eq:cumulants1}
    K_1 &=& \mu=\langle N\rangle \\
    K_2 &=& \sigma^2=\langle N^2\rangle-\langle N\rangle^2 \\
    K_3 &=& S\sigma^3=\langle (N-\langle N\rangle)^3\rangle
\end{eqnarray}

where $\mu$ is event averaged particle mean expectation value, $S$ is the skewness and $\sigma^2$ is variance. The $\langle\cdot\rangle$ notation implies the event-average of the number of baryons (or any other specific particle number) $N$.

Instead of using the cumulants, which are to first order proportional to the system volume, one often uses cumulant ratios to remove this explicit volume dependence:
\begin{eqnarray}
\label{eq:cumulants2}
    \frac{K_2}{K_1}&=&\frac{\sigma^2}{\mu} \label{K2}\\
    \frac{K_3}{K_2}&=&S\sigma \label{K3}
\end{eqnarray}

The observable cumulants are sensitive to many aspects of the collision dynamics, not only the presence of a phase transition. Three effects which are of particular importance are the fluctuations of the number of participants \cite{Skokov:2012ds}, proper treatment of the experimental acceptance, and corrections due to experimental efficiency \cite{Luo:2018ofd}. In our study we will essentially neglect these effects as a proper treatment would require detailed knowledge on a specific detector system and centrality determination method as well as the corresponding corrections. We simply assume that any signal surviving, even though we include the effects of volume fluctuations etc., should also be visible in the experimental data. On the other hand, any signal that becomes only visible after significant (sometimes model dependent) corrections may not be very reliable. 

\subsection{Cluster production}

In the UrQMD transport model, the production of nuclear clusters is implemented using phase-space coalescence. Nucleons at kinetic freeze-out that are sufficiently close in both coordinate and momentum space combine into composite particles. The procedure operates in several steps starting with the protons and neutrons obtained from UrQMD at kinetic freeze-out. To check whether a number of nucleons may form a light nucleus, one then transforms in the local rest frame of the nucleons and checks if the relative momentum $\Delta p < p_{max}$ and relative distance $\Delta r < r_{max}$ in that frame is smaller than the coalescence parameters $p_C$ and $r_C$. These parameters have to be defined for each nucleus, a more detailed description of the procedure can be found in \cite{Sombun:2018yqh,Hillmann:2021zgj}. When using the UrQMD model with a density dependent potential, the interactions cause additional correlations between the nucleons at freeze-out. This can also lead to effective bound states as discussed in \cite{Kireyeu:2023spj,Kireyeu:2022qmv}.

\begin{table}[b]
\centering
\begin{tabular}{|c|c|c|c|c|c|c|c|}
\hline
 & d & t & $^3$He & $^4$He \\
\hline
spin-isospin projection & 3/8 & 1/12 & 1/12 & 1/96 \\
\hline
$\Delta r_{max}$ [fm] & 4.0 & 3.5 & 3.5 & 3.5 \\
\hline
$\Delta p_{max}$  [GeV] & 0.3 & 0.45 & 0.45 & 0.55 \\
\hline
\end{tabular}
\caption{Parameters used in the UrQMD  phase-space coalescence when potentials are included.\label{tab:fac}}
\end{table}

\begin{figure*}[t]
  \centering
        \includegraphics[width=0.49\textwidth]{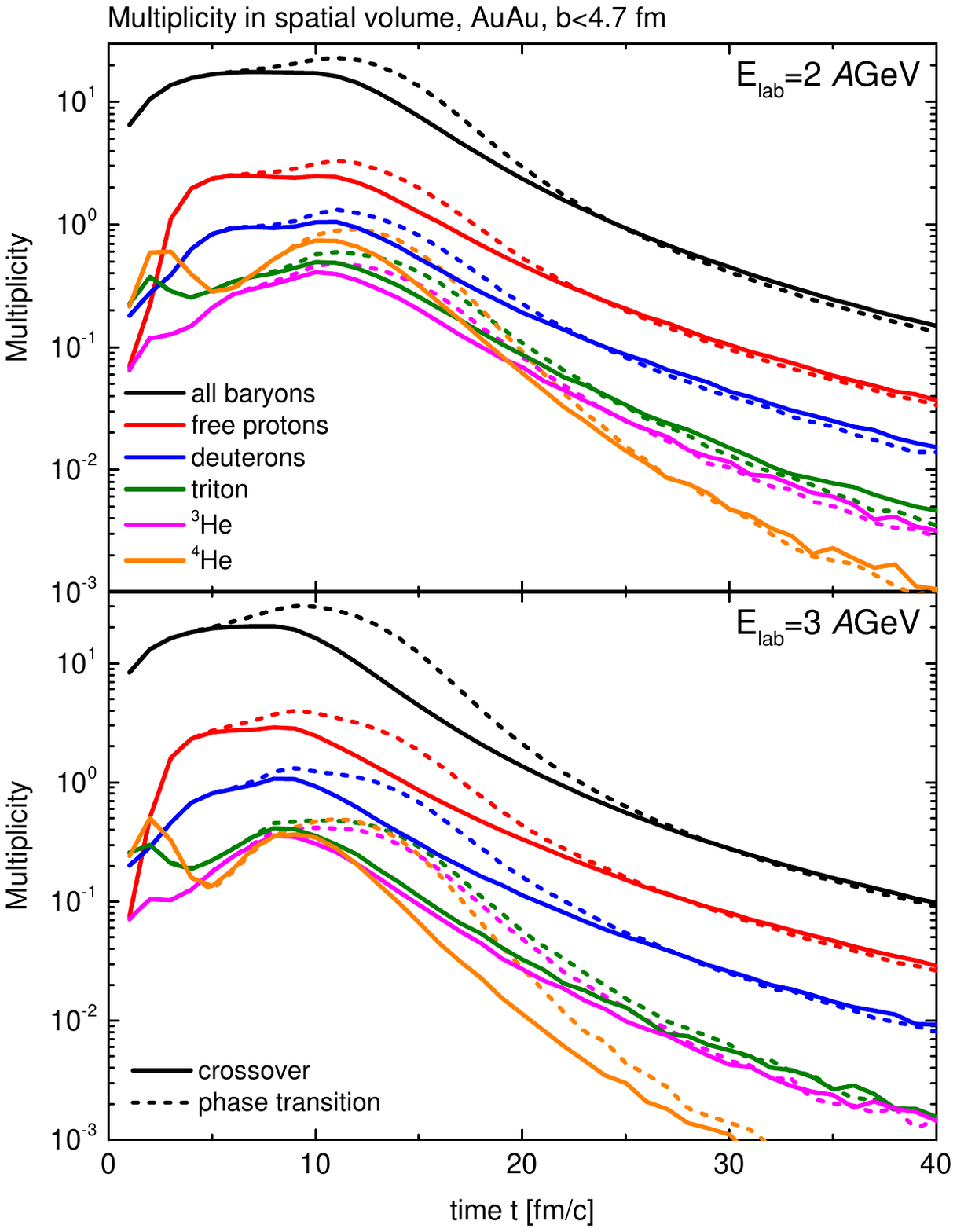}
        \includegraphics[width=0.49\textwidth]{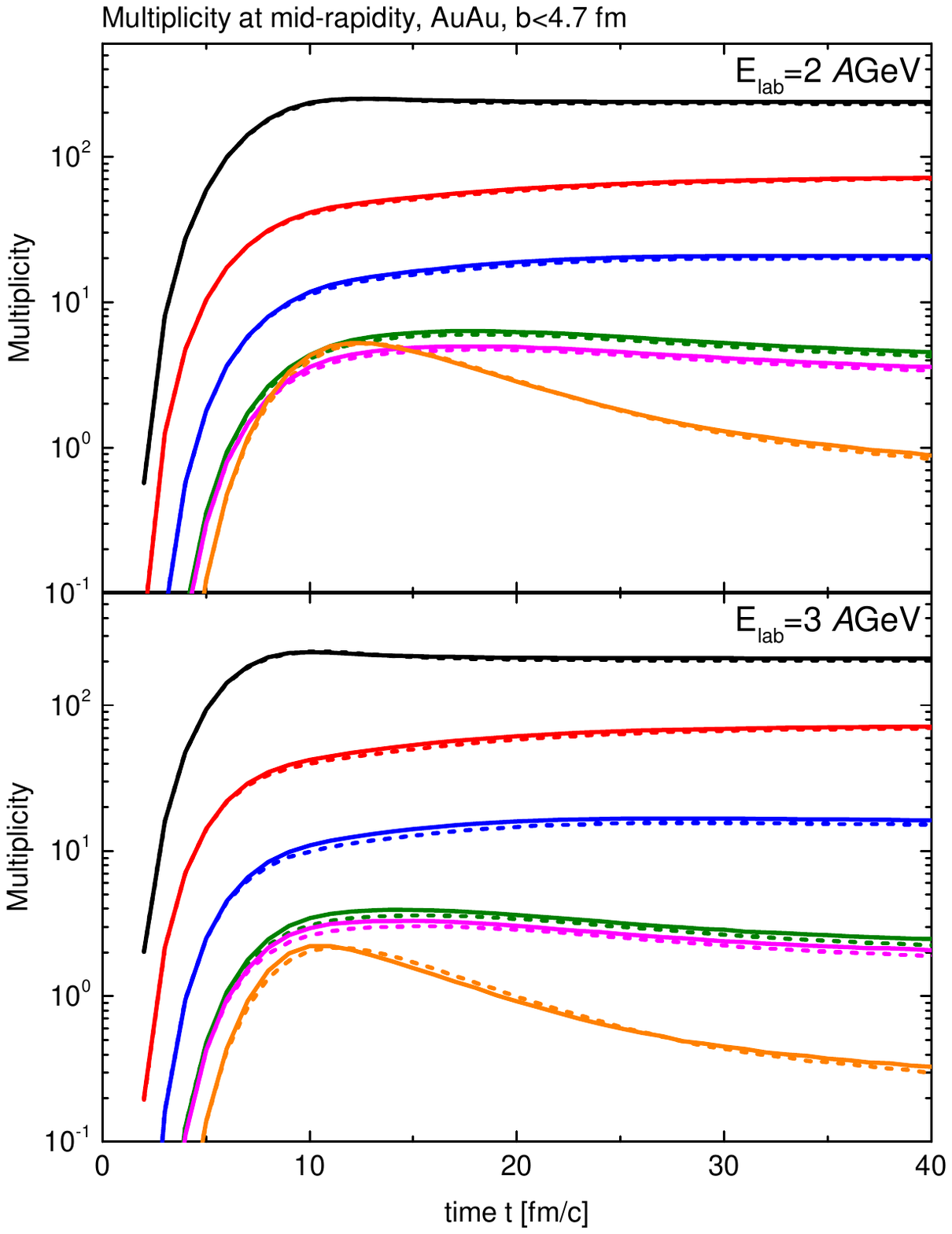}
  \caption{(Color online) Time evolution of the multiplicities of light nuclei ($p$ (red), $d$ (purple), $t$ (green), $^3He$ (orange), $^4He$ (blue) and all baryons (grey)) at $E_{lab}=2A$ (upper panels) and $3A$\,GeV (lower panels). The left panels show the multiplicities within the spherical spatial volume with radius 2\,fm centered around the origin from UrQMD calculations with a CMF EoS with phase transition (dashed lines) and with a crossover (solid lines). The right panels represent the multiplicities at mid-rapidity $|y|\leq0.5$ from the same calculations. Enhancements can be seen for both energies in a spatial volume for the scenario with a phase transition.} 
  \label{fig:multi}
\end{figure*}

It is therefore necessary to refit the coalescence parameters of all the light nuclei. For the deuteron and triton we have used the same data as in \cite{Sombun:2018yqh,Hillmann:2021zgj} while for the $^4$He we have used data from the E864 experiment \cite{E864:2000auv}. The resulting fit parameters are independent of collision energy and are summarized in table \ref{tab:fac}. The approach has provided a reliable description of available experimental data over a wide range of energies and system sizes.

\section{Results}

As shown in previous studies, the inclusion of a first-order phase transition in the simulations of heavy ion reactions in UrQMD will lead to enhanced fluctuations of the baryon number in coordinate space \cite{Savchuk:2022msa,Kuznietsov:2023iyu} and therefore a deviation of the baryon number susceptibilities from their baseline. In the following, we will extend this study to the third order susceptibility and show the effect of clustering in coordinate and momentum space. To do so, we have selected two specific beam energies, $E_{\mathrm{lab}}=2$ and $3A$\,GeV for which $64000$ and $80000$ central AuAu collisions ($b<4.7$\,fm), respectively, have been simulated. With this specific choice of beam energies and the constructed phase transition, the system at $E_{\mathrm{lab}}=2A$\,GeV will just enter the mixed phase during compression from below and the system created at the beam energy of $E_{\mathrm{lab}}=3A$\,GeV will slightly overshoot the mixed phase and enter it from above, leading to a different time dependence of the susceptibilities. Note that this is of course only true for our specific choice of coexistence densities and not a general statement about the beam energy related to the actual occurrence of such a phase transition.

\begin{figure*}[t]
    \centering
        \includegraphics[width=0.49\textwidth]{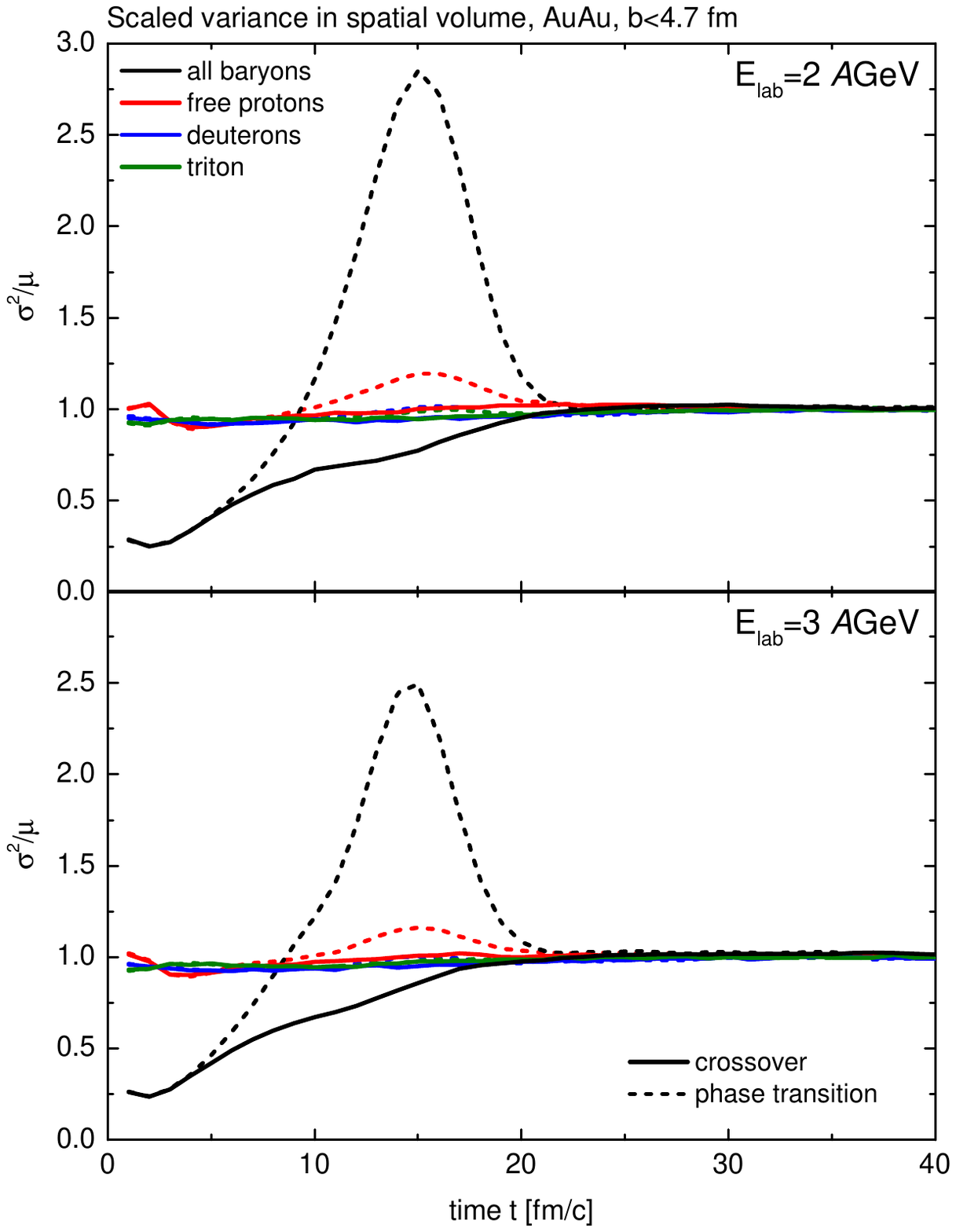}
        \includegraphics[width=0.49\textwidth]{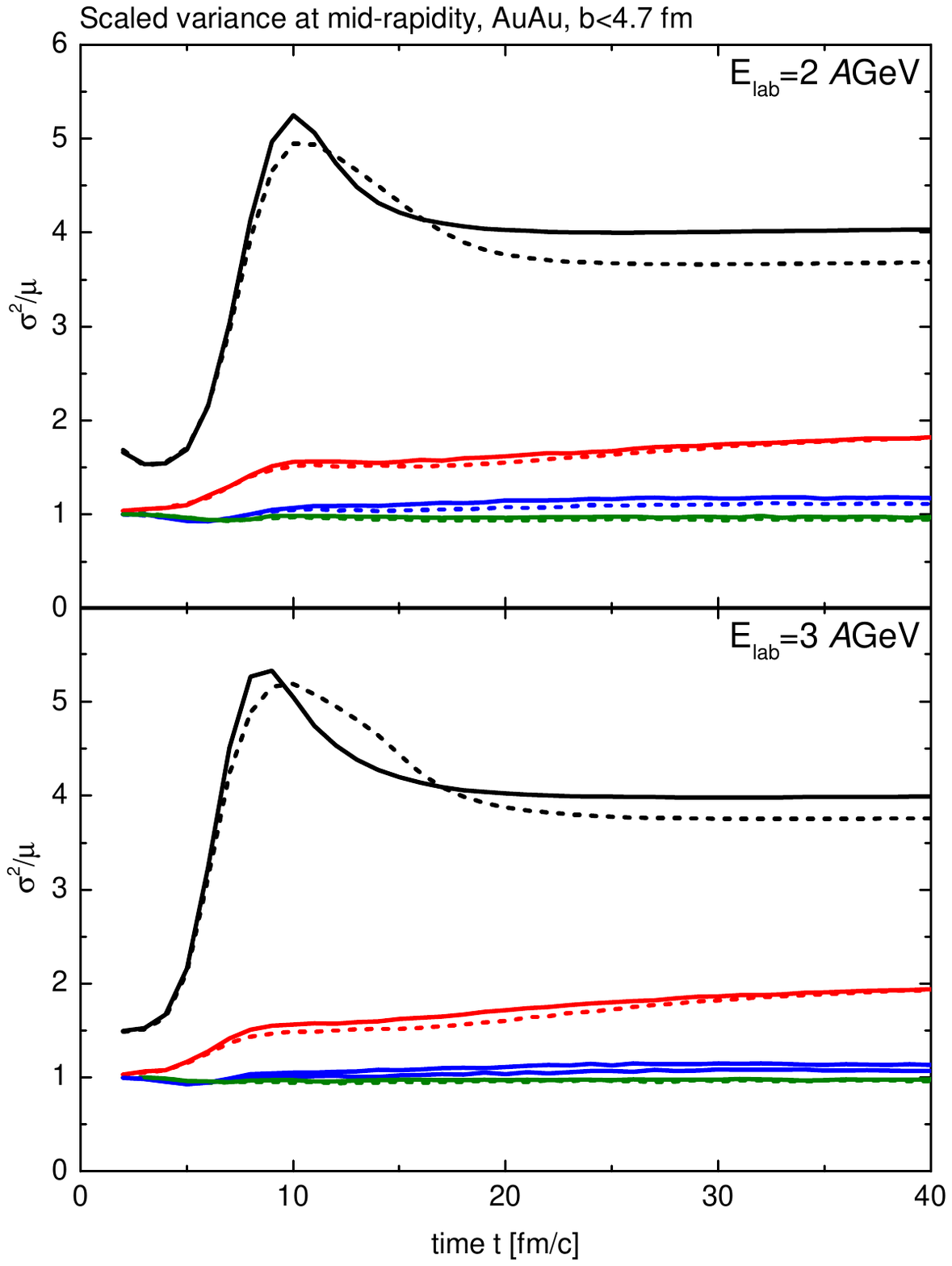}
    \caption{\label{fig:var1}(Color online) Time evolution of the scaled variance of the multiplicities of baryons (black), protons (red) and light nuclei (deuterons (blue), tritons (green)) at $E_{lab}=2A$\,GeV (upper panels) and $3A$\,GeV (lower panels). The left plots show the results within a spatial volume and the right figures correspond to results within the rapidity range $|y|\leq0.50$. The UrQMD calculations were performed with a CMF EoS with a phase transition (dashed lines) and with a crossover (solid lines). During the phase transition, an explicit enhancement in the baryon cumulants can be seen while the free protons and other nuclear clusters are only mildly influenced. 
    }
\end{figure*}

\subsection{Multiplicities}

The left two panels of Fig.~\ref{fig:multi} show the time evolution of several particles and clusters such as the baryon number, the number of free protons as well as the multiplicity of light nuclei up to Helium-4 in a fixed central volume of radius $2$\,fm around the center of the collision in the center of mass frame of the collisions at $2$ and $3A$\,GeV. The light nuclei multiplicities at a given time $t$ are calculated under the assumption that no rescattering would take place after that, which is of course incorrect, but provides the approximate yield of light nuclei at any given time assuming the system would decouple at that instant. While the solid lines represent the multiplicities for the default crossover CMF EoS, the dashed lines correspond to the EoS with a phase transition. At the early times and late times the multiplicities for both equations of state are very similar. As the system enters the coexistence phase, between around $8$ to $24$\,fm/c, the average baryon density in the phase transition scenario is increased relative to the crossover which leads to an increase in baryon number and all related multiplicities in the same central spatial volume. 

Note that the high net-baryon density present in the early stage of the collision favors the relatively compact cluster He-4 compared to triton and He-3. After an initial increase in all particle types, one observes a decrease at later times as these leave the volume. In contrast to that, the multiplicities in momentum space, shown on the right panels, saturate after an initial increase up to around $8$\,fm/c, with the exception of He-4 that undergoes a significant decrease. This is a result of the decreasing density during later times of the evolution which makes the production of this more strongly bound cluster more and more unlikely. 

The evolution in momentum space overall shows no significant difference between the crossover and the phase transition scenarios, the yields are nearly the same for all investigated particle species. This means that the phase separation effects only the coordinate space configuration of the baryons in the dense phase and has little effect on the momentum space fluctuations.

\subsection{Cumulant ratios}

\begin{figure*}[t]
    \centering
        \includegraphics[width=0.49\textwidth]{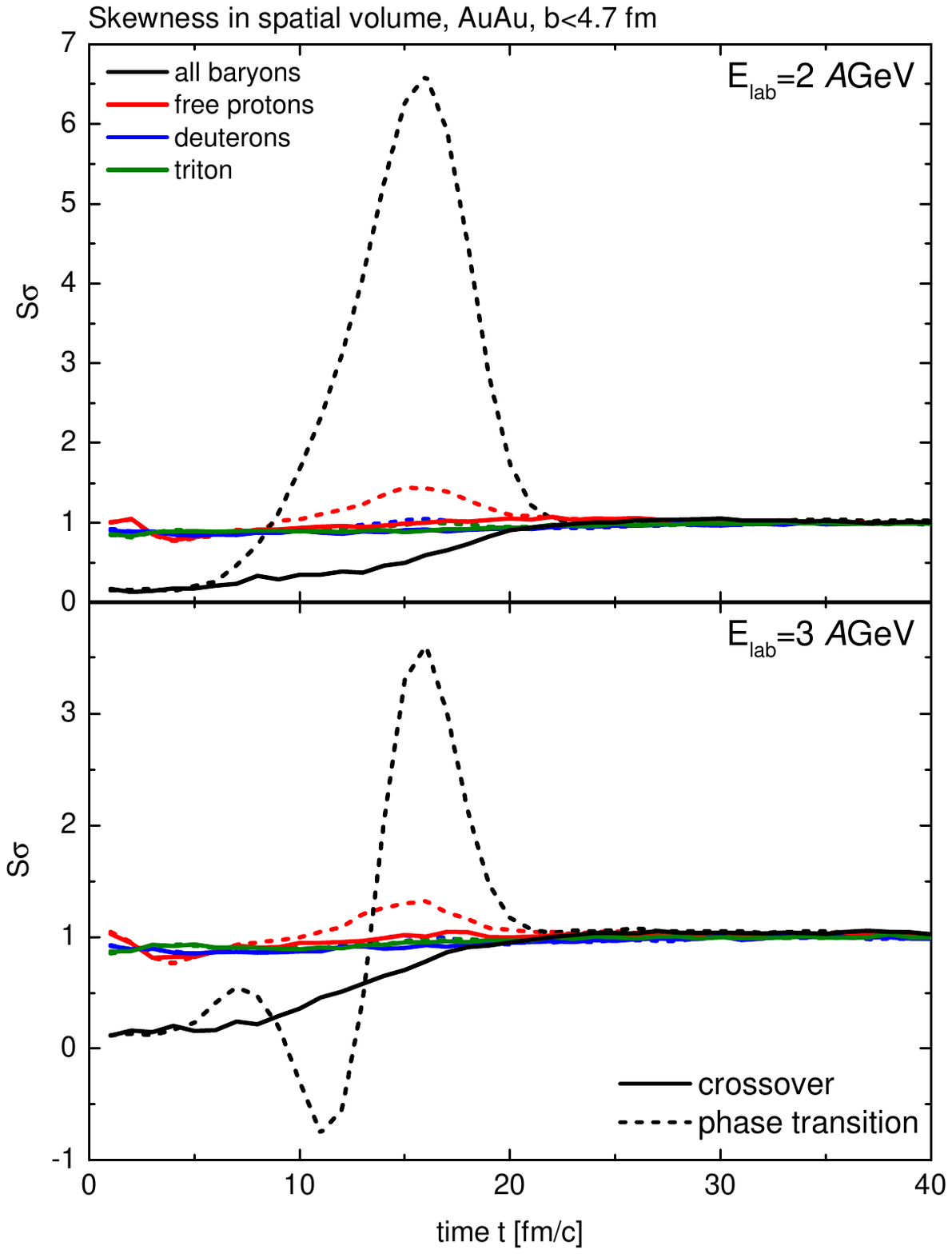}
        \includegraphics[width=0.49\textwidth]{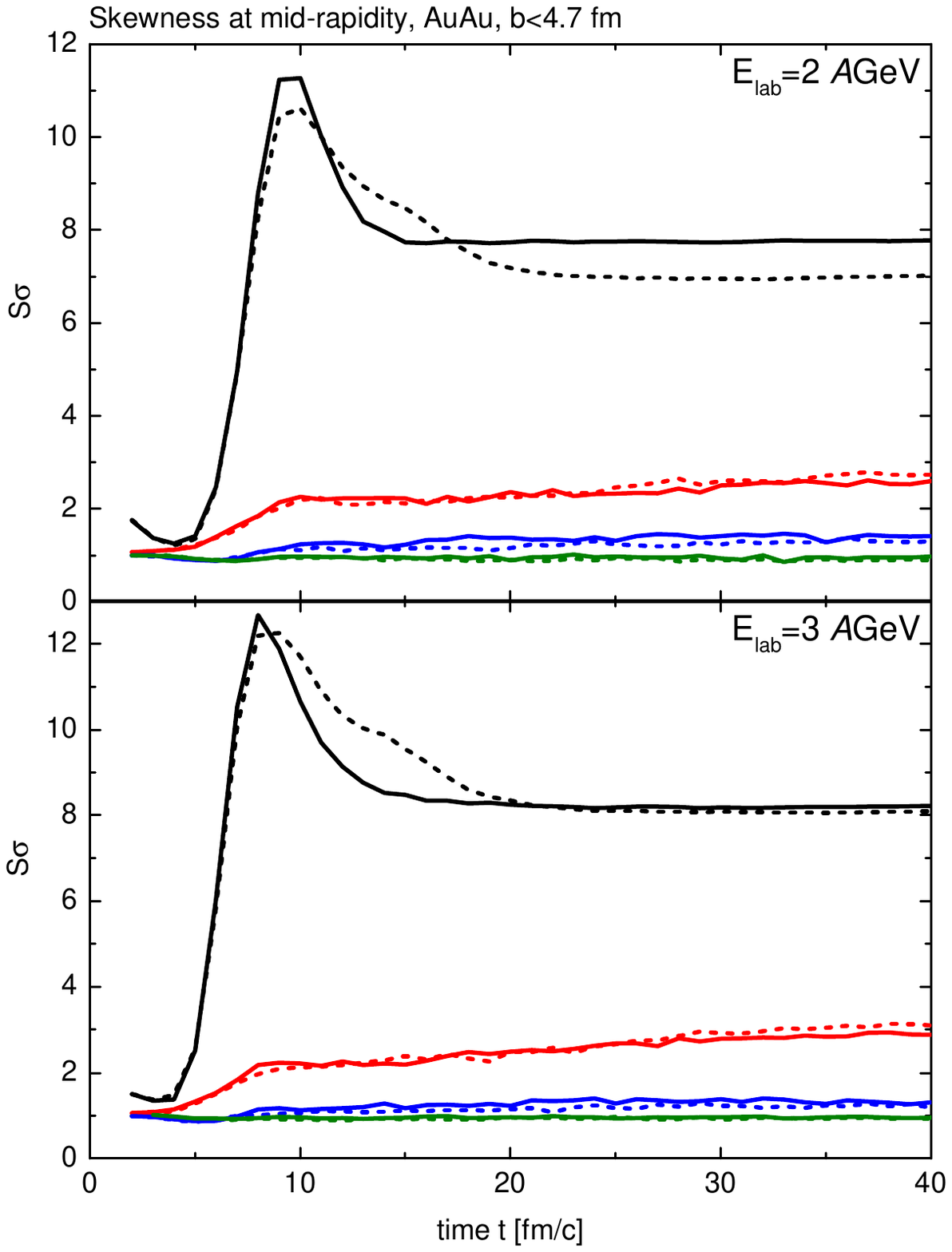}
    \caption{(Color online) Time evolution of the skewness of the multiplicities of baryons (black), protons (red) and light nuclei (deuterons (blue), tritons (green)) at $E_{lab}=2A$\,GeV (upper panels) and $3A$\,GeV (lower panels). The left plots show the results within a spatial volume and the right figures correspond to results within the rapidity range $|y|\leq0.50$. The UrQMD calculations were performed with a CMF EoS with a phase transition (dashed lines) and with a crossover (solid lines).}
    \label{fig:var2}
\end{figure*}

It is now interesting to see whether the presence of a phase transition will influence the baryon number fluctuations, and therefore correlations, in phase space.
Therefore, we turn to the time evolution of cumulant ratios according to Eq.~\eqref{eq:cumulants1} starting with the scaled variance in Fig.~\ref{fig:var1} at $E_{lab}=2$ and $3A$\,GeV (upper and lower panels). In coordinate space, we observe that due to the conservation of the total baryon number, the values for baryons start below the Poisson base value of $1$ in contrast to the other particles. We find a significant enhancement in $\frac{\sigma^2}{\mu}$ for a phase transition compared to a crossover, which peaks at the time of highest compression. A less significant enhancement is observed for protons, as they make up only a small fraction of the total baryon number due to resonance excitations as well as the presence of a large number of clusters. Nuclear clusters experience an even smaller, yet still discernible enhancement. 

In momentum space, cf. the right column of Fig.~\ref{fig:var1}, we find that there is a separation in the curves for the crossover and phase transition scenario for all baryons after around $8$\,fm/c. A previous work discussed the relation of coordinate and momentum space cumulants in a box simulation \cite{Kuznietsov:2024xyn} and found that the strong enhancement of fluctuations in the coordinate space is not translated into enhanced cumulant ratios in momentum space. In our work we can study this relation in a fully dynamic scenario. We also find that the strong enhancement in the cumulant ratios in coordinate space is not reflected in the results in momentum space. More precisely, for a phase transition, the peak in the scaled variance appears slightly later and is broader than for a crossover. Interestingly, the asymptotic (observable) value appears to be suppressed compared to a crossover evolution. This can be attributed to a harder equation of state in the latter case creating a stronger density gradient and thus pressure at early times which reduces the number of spectators and thus leads to more fluctuations in the number of participants (or volume). This effect is also visible in the investigated light nuclei. 

The situation is similar for $S\sigma$, see Fig.~\ref{fig:var2}. An obvious difference can be seen at the higher investigated energy of $3A$\,GeV, see the lower left panel. Here, the phase transition leads to an enhancement, followed by a suppression and another enhancement of the net-baryon number cumulant compared to the crossover. This is a result of the qualitatively different evolutions for the two energies, where only for $3A$\,GeV the system overshoots the spinodal regions and crosses two spinodal lines during the expansion.

\begin{figure*}[t]
    \centering
    \includegraphics[width=0.49\textwidth]{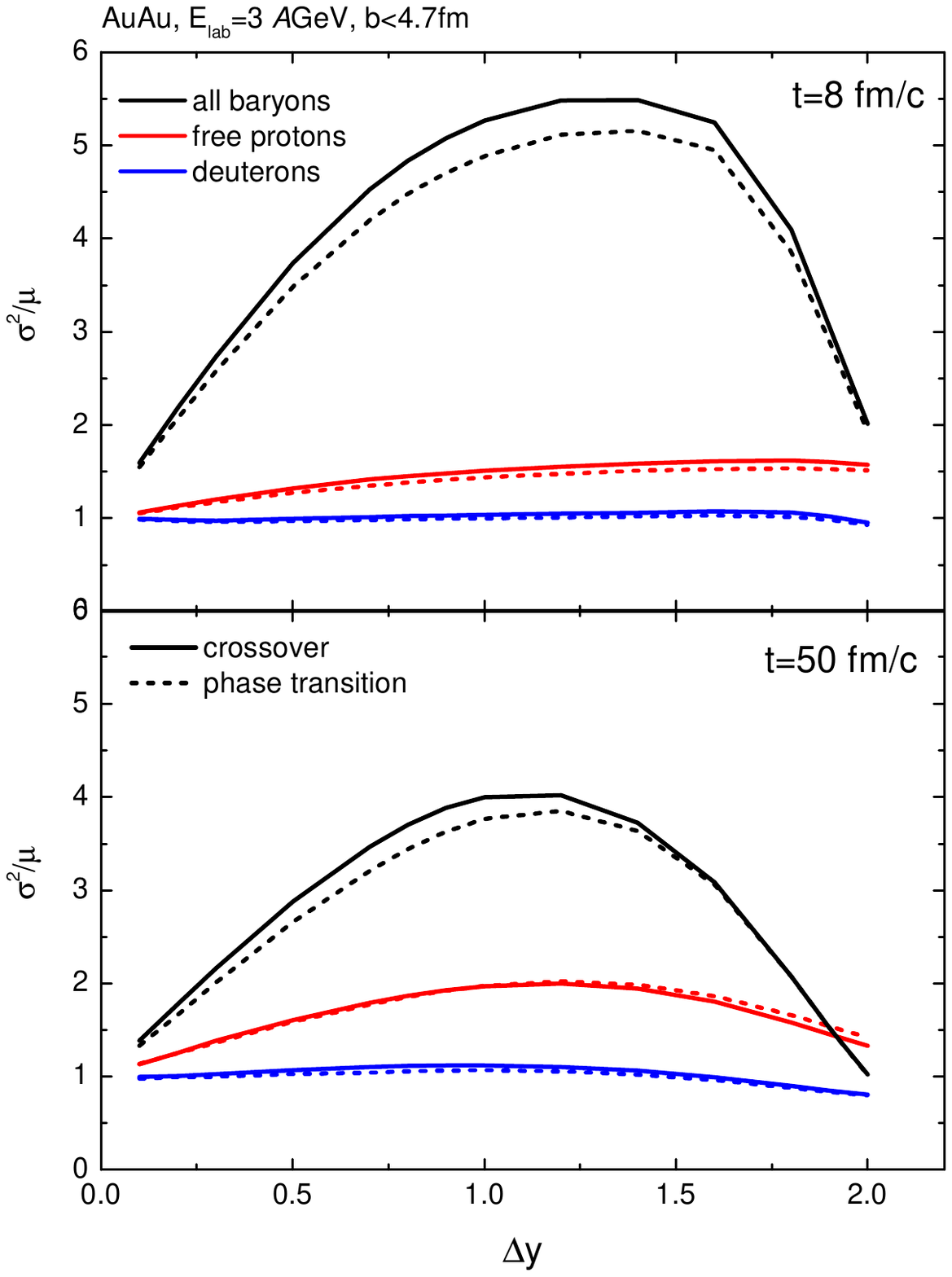}
    \includegraphics[width=0.49\textwidth]{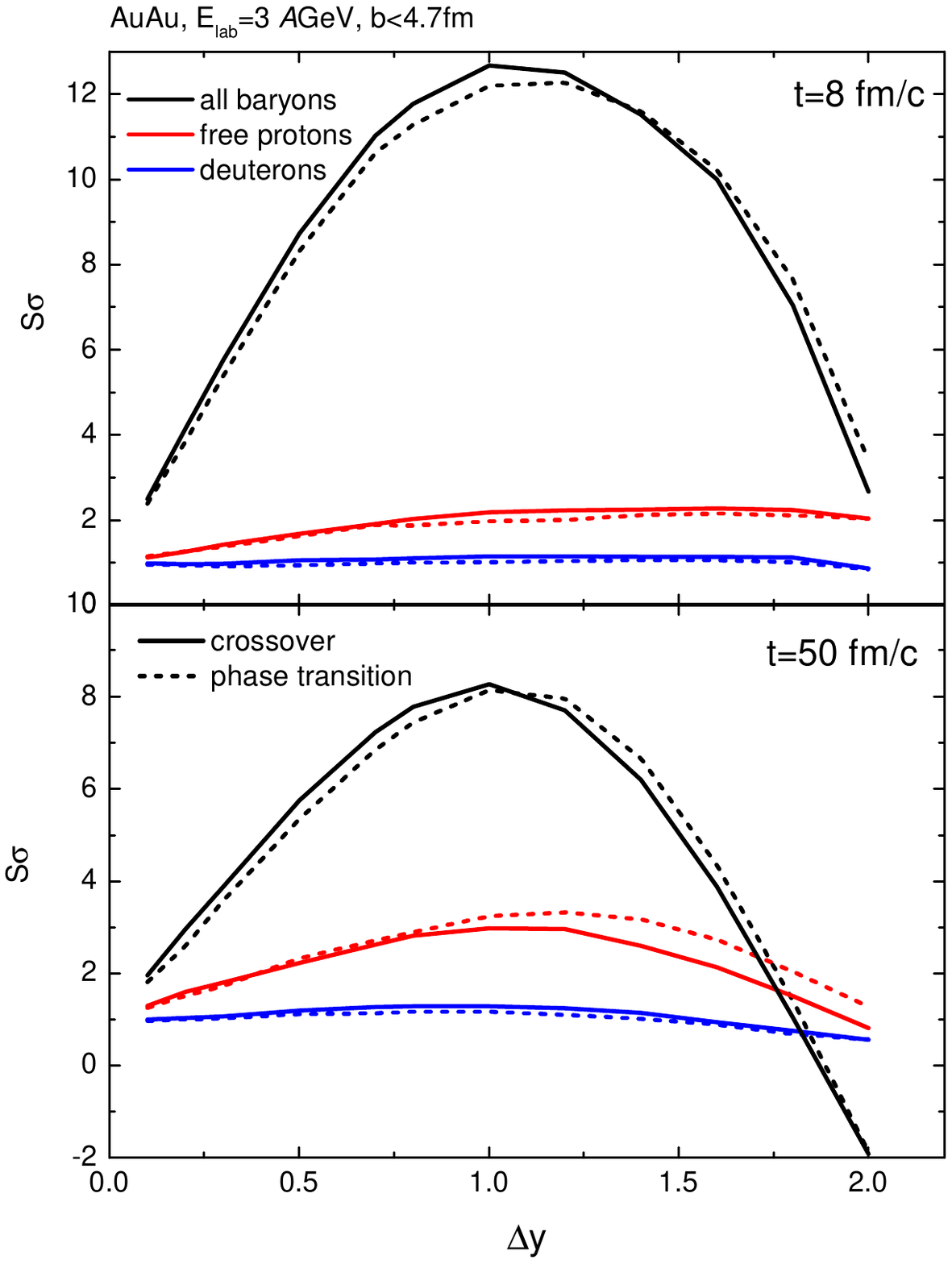}
    \caption{(Color online) Dependence of the scaled variance (left panels) and skewness (right panels) on the size of the rapidity acceptance $\Delta y$. The upper plots show the dependence at a fixed time of $8$\,fm/c and the lower at $50$\,fm/c. Shown are results for all baryons (black lines), free protons (red lines) and deuterons (blue lines) at $E_{lab}=3A$\,GeV from UrQMD calculations with a CMF EoS with a phase transition (dashed lines) and without a phase transition (solid lines).}
    \label{fig:rap}
\end{figure*}

Similar to the case for the scaled variance, the evolution in momentum space reveals no clear distinguishing signal between the two scenarios. While the suppression of $S\sigma$ at late times is also present for $2A$\,GeV, it is far less significant at $3A$\,GeV, which is consistent with findings that volume fluctuations are more important for the second order cumulants than the third order \cite{Bzdak:2016jxo}.

\subsection{Cumulants in different rapidity windows}

Even though we do not observe any significant signal in the cumulants on momentum space, it is nevertheless important to check which choice of acceptance window in rapidity would be best suited experimentally to study the cumulant rations.
We therefore show the dependence of the cumulants for the number of baryons, deuterons, and free protons on the width of the rapidity window in Fig.~\ref{fig:rap}. The rapidity window is defined by $\Delta y=y_{max}-y_{min}$ where $y_{max}$ and $y_{min}$ are the maximum rapidity and minimum rapidity, respectively, with $y_{max}=-y_{min}$, and $y_{max}>0$. We have chosen rapidity windows at two fixed time steps of $8$\,fm/c (left) and $50$\,fm/c (right) to compare fluctuations in the initial and late stage of the collision. The largest baryon fluctuations are observed around $\Delta y=1.0$, for very small windows the fluctuations become Poissonian and for large ones, baryon number conservation plays the dominant role. Overall, as one expects, there are less fluctuations in the proton number compared to the baryon number. 

While for the scaled variance the effects of the participant fluctuations dominate for all rapidity windows, the skewness of free protons does show a slight enhancement in case of a phase transition for larger rapidity windows. For the beam energies presented here we therefore suggest that the skewness in a rapidity window of $\Delta y > 1$ is best suited to observe any signals of a phase transition. 

\begin{figure*}[t]
    \centering
    \includegraphics[width=0.49\textwidth]{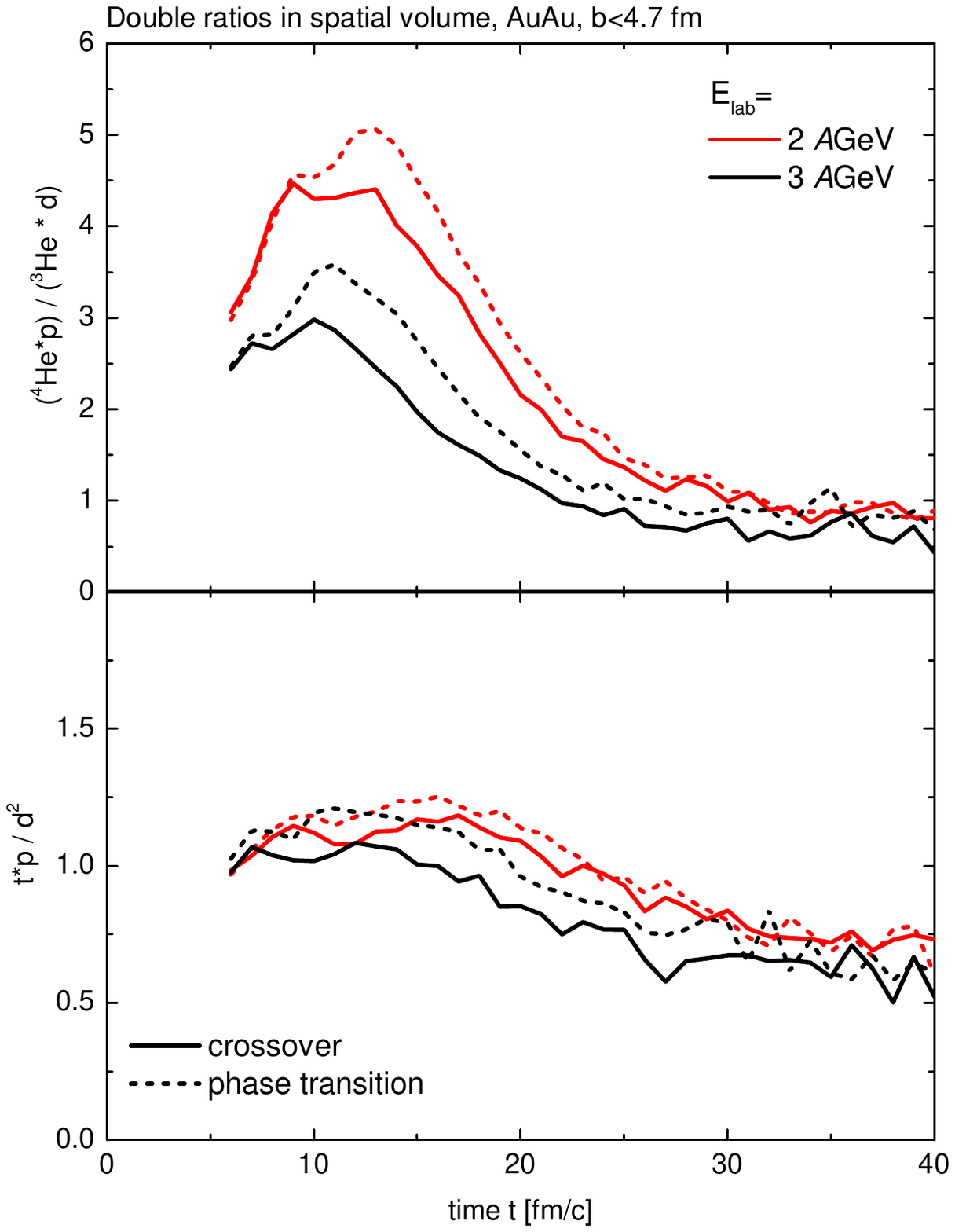}
    \includegraphics[width=0.49\textwidth]{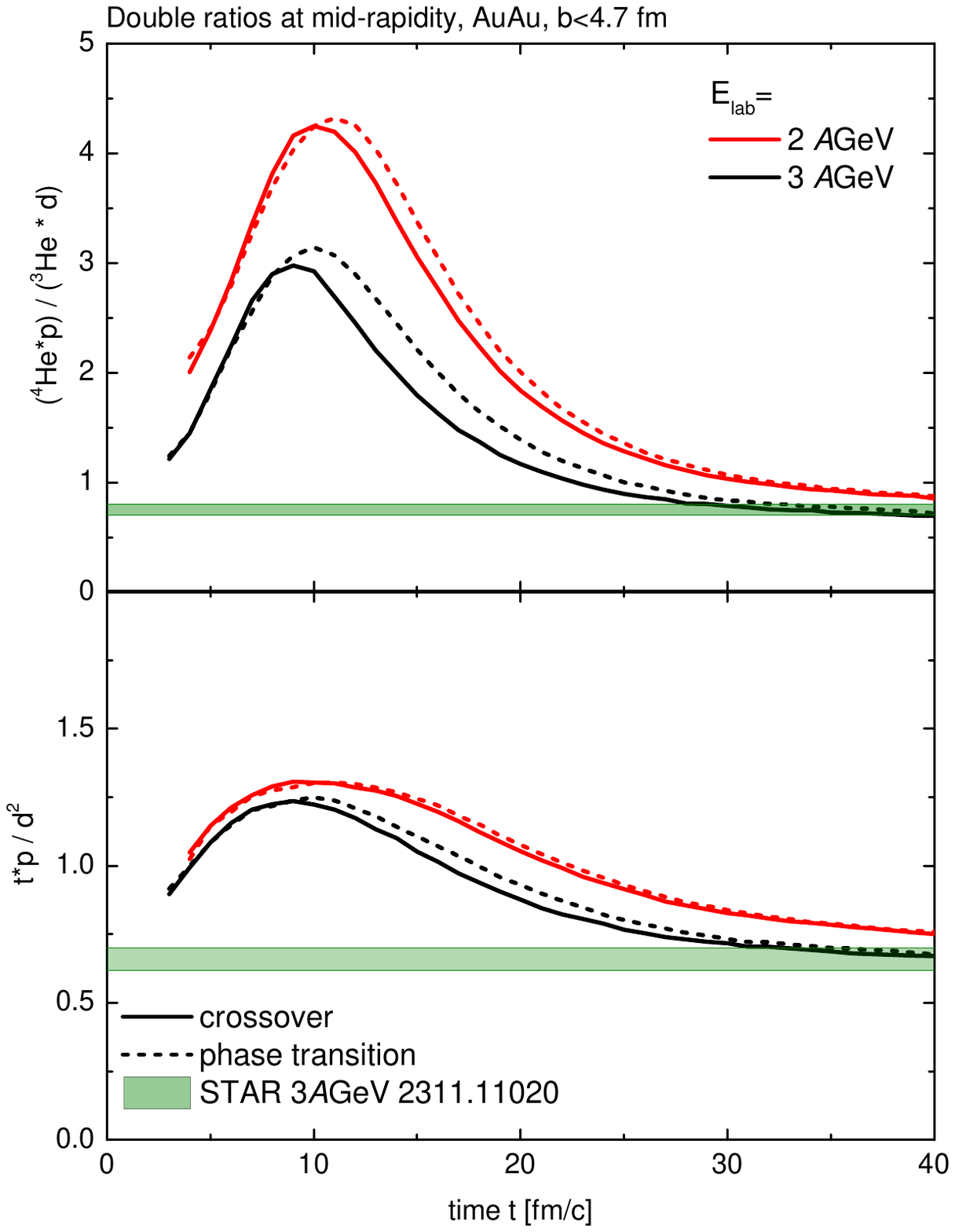}
    \caption{(Color online) Time evolution of double ratios $(N_{^{4}He} \times N_{p})/(N_{^{3}He} \times N_{d})$ (upper panels) and $N_{t} \times N_{p}/N_{d}^2$ (lower panels) for corresponding beam energies of $2$ and $3A$\,GeV in the spatial volume (left) and within a rapidity window $|y|\leq0.5$ (right) with phase transition (dashed lines) and with crossover (solid lines). In all cases, an enhancement with a phase transition becomes visible from around $10$\,fm/c on. The green bands correspond to measurements from the STAR experiment at $E_{lab}=3A$\,GeV \cite{STAR:2023uxk}.}
    \label{fig:ratios}
\end{figure*}

\subsection{Light nuclei ratios}

After confirming that the phase transition will indeed lead to fluctuations of the baryon number in coordinate space we now investigate if this will lead to significant effects in the production of small nuclear clusters. From a simplified coalescence picture one can argue that the enhanced fluctuations in coordinate space will lead to an enhancement of certain double ratios of nuclei \cite{Sun:2020uoj}. However, it is still disputed whether this will also lead to a significant observable enhancement of these ratios in the final state, as the fluctuations in phase space rapidly disappear in the late stage and the final size of the system will lead to local conservation effects reducing the number of nucleons available for cluster formation in a local region of phase space.
In the following we will focus on the time evolution of the double ratios $(N_{^{4}He} \times N_{p})/(N_{^{3}He} \times N_{d})$ and $N_{t} \times N_{p}/N_{d}^2$. Fig.~\ref{fig:ratios} shows the time evolution of the double ratios for both energies in coordinate space (left) and momentum space (right). We see that for all curves the ratios start to increase until around $10$\,fm/c, when a slight enhancement in the phase transition sets in compared to the crossover. Interestingly, this enhancement is also clearly visible in the ratio $(N_{^{4}He} \times N_{p})/(N_{^{3}He} \times N_{d})$ extracted in the rapidity window, for both investigated beam energies. This is different to the cumulant ratio which never showed a clear signal in momentum space. However, after a sufficiently long time, all ratios approach the same value. For completeness we also compare the double ratios to measurements from the STAR experiment at $E_{lab}=3A$\,GeV \cite{STAR:2023uxk} which is in good agreement with the values at late times from our simulations for both equations of state.

\section{Discussion} 

In summary, we have investigated the impact of a first-order phase transition on baryon and light nuclei yields, ratios, and cumulants at low beam energies using the UrQMD transport model with a modified equation of state. We have studied the time evolution of these observables both in coordinate and in momentum space, the latter being relevant for experiments. 

Overall, our results in a central spherical spatial volume have all shown a clear modification of the respective observable in the presence of a strong first-order phase transition compared to a crossover. The strongest effect was seen for the cumulant ratios $\sigma^2/\mu$ and $S\sigma$ of the net-baryon number.

For the experimentally accessible proton number cumulants, the most promising observable seems to be the higher order cumulants for free protons at larger rapidity windows $\Delta y>1$.  

Within a rapidity window of $\Delta y=1$, the only relatively clear distinguishable and potentially useful effect in nuclei production was found for the light nuclei ratio $(N_{^{4}He} \times N_{p})/(N_{^{3}He} \times N_{d})$, where a roughly $10$\% larger signal is found during intermediate times. 
However, at late times, the signal disappears, making it unlikely that this effect is measurable. 

Future studies should address a more realistic and 'closer to experiment' determination of the cumulants including corrections for volume fluctuations and experimental acceptance and efficiency.

\begin{acknowledgments}
T.R. acknowledges support through the Main-Campus-Doctus fellowship provided by the Stiftung Polytechnische Gesellschaft Frankfurt am Main (SPTG). 
The authors acknowledge for the support the European Union's Horizon 2020 research and innovation program under grant agreement No 824093 (STRONG-2020). This research has received funding support from the NSRF via the Program Management Unit for Human Resources and Institutional Development, Research and Innovation [Grant No. B16F640076]. This work was supported by the DAAD (PPP Thailand).
The computational resources for this project were provided by the Center for Scientific Computing of the GU Frankfurt and the Goethe-HLR.
\end{acknowledgments}

\bibliography{refs}

\end{document}